# Bioabsorbable WE43 Mg alloy wires modified by continuous plasma electrolytic oxidation for implant applications. Part II: degradation and biological performance

Wahaaj Ali[1,2,3], Mónica Echeverry-Rendón[1], Guillermo Dominguez[1,4], Kerstin van Gaalen[3,5], Alexander Kopp[3], Carlos González[1,4], and Javier LLorca[1,4*]

[1]IMDEA Materials, C/Eric Kandel 2, 28906 Getafe, Madrid, Spain
[2]Departament of Material Science and Engineering, Universidad Carlos III de Madrid, Leganés, Madrid 28911, Spain
[3]Meotec GmbH, Philipsstr. 8, 52068 Aachen, Germany
[4]Department of Materials Science, Polytechnic University of Madrid/Universidad Politécnica de Madrid, 28040 Madrid, Spain
[5]Biomechanics Research Centre (BioMEC), Biomedical Engineering, School of Engineering, University of Galway, Galway, Ireland

**Abstract**

The corrosion, mechanical degradation and biological performance of cold-drawn WE43 Mg wires were analyzed as a function of thermo-mechanical processing and the presence of a protective oxide layer created by continuous plasma electrolytic oxidation (PEO). It was found that the corrosion properties of the non-surface-treated wire could be optimized by means of thermal treatment within certain limits, but the corrosion rate remained very high. Hence, strength and ductility of these wires vanished after 24 h of immersion in simulated body fluid at 37ºC and, as a result of that rather quick degradation, direct tests did not show any MC3T3-E1 preosteoblast cell attachment on the surface of the Mg wires. In contrast, surface modification of the annealed WE43 Mg wires by a continuous PEO process led to the formation of a homogeneous oxide layer of $\approx$ 8 µm and significantly improved the corrosion resistance and hence the biocompatibility of the WE43 Mg wires. It was found that a dense layer of Ca/P was formed at the early stages of degradation on top of the $Mg(OH)_2$ layer and hindered the diffusion of the $Cl^-$ ions which dissolve $Mg(OH)_2$ and accelerate the corrosion of Mg alloys. As a result, pitting corrosion was suppressed and the strength of the Mg wires was above 100 MPa after 96 h of immersion in simulated body fluid at 37ºC. Moreover, many cells were able to attach on the surface of the PEO surface-modified wires during cell culture testing. These results demonstrate the potential of thin Mg wires surface-modified by continuous PEO in terms of mechanical, degradation and biological performance for bioabsorbable wire-based devices.

## 1. Introduction

Application of Mg wires for ligature of bleeding vessels was first reported by Huse in 1878 but use of Mg for biomedical applications in implants has not been reported until the 21st century because of the concerns associated with the fast degradation of Mg [1]. In the last decade, Mg alloys-based devices have been tested to fix bone fractures or treat vessel stenosis mainly in Germany, but also China, and Korea. Initiated by companies such as Syntellix GmbH (Hannover, Germany), BIOTRONIK SE & Co. KG (Berlin, Germany) and recently Medical Magnesium GmbH (Aachen, Germany), Mg-Y-Re-(Zr)-based stents and screws achieved approval for use in the European medical device market [2].

Seitz et al. [3][4] probably were the first to rejuvenate also the idea of bioresorbable Mg wires to replace polymer sutures for medical applications. Afterward, numerous studies have reported the manufacturing and characterization of Mg alloy wires for different bioabsorbable (or bioresorbable or biodegradable) applications including pins for orthopedic fracture fixation [5], electrodes for neural recording [6] and cardiac pacemakers [7], staples and sutures for wound closure [8], cardiovascular [9] and biliary [10] stents, etc. However, to our knowledge, medical devices using Mg in form of wires have not been commercially approved so far, even though Mg alloy wires present excellent mechanical properties (elastic modulus of 40-45 GPa and tensile strength in the range 100-600 MPa [11]) . Obviously, the main concern for the application of thin Mg wires in the medical field, in particular with diameters < 1 mm, is the fast corrosion rate which can lead to a significant reduction of mechanical properties and disintegration of wires in short time [12]. This process is accelerated by the small grain size (usually < 10 µm) of Mg wires manufactured by cold drawing [12][13] and by the high specific surface of such wires [11] [14]. Moreover, corrosion can be accelerated during *in vivo* degradation if the wires are subjected to tensile stresses [5][13][15]. Surface reactivity of the wires can be controlled up to some extent by means of annealing treatments [11][16][12][17], but often full degradation of the wires will be found before full healing can be achieved in an *in vivo* setup. Chemically, the high reactivity of Mg is associated with a local increase in pH and the release of hydrogen gas which do not favor cytocompatibility [18]. Therefore, a cost-effective and reliable strategy is necessary to slow down corrosion of thin Mg wires in biological environments and to pave their way to the medical device market and enable innovative bioabsorbable implant concepts.

There is ample experimental evidence in the literature showing that surface-modification by plasma electrolytic oxidation (PEO) controls very effectively the rash degradation of Mg and its alloys. For instance, Rendenbach et al. [19] demonstrated excellent *in vitro* and *in vivo* biocompatibility of Mg plates and screws from a Mg-Y-RE-(Zr) alloy (WE43MEO) surface modified by PEO using Kermasorb$^©$ electrolyte (Meotec GmbH, Aachen). The implants successfully converted into bone, and no toxic or even somehow adverse response was observed in the surrounding tissue. Good cytocompatibility was also reported by Li et al. [15] for porous WE43 Mg scaffolds manufactured by laser power-bed fusion at least when the surface was PEO modified, while cell response tended to be less favorable in the absence of the PEO oxide layer. It also showed that the barrier and biological properties of the oxide layers obtained by PEO can be adjusted to more specific requirements by controlling electrical parameters and electrolyte composition [20][21].

Application of PEO to modify specifically the surface of Mg wires was first reported by Chu et al. [10]. However, it was only done on a small segment of AZ31 Mg wire which acted as anode in the electrolyte bath during PEO. Obviously, this strategy was not suitable to process and treat by PEO continuous wires that could be used in medical devices. Hence, in the first part of our work [22], we have reported a continuous process to modify the surface of Mg wires of unlimited length by means of PEO. The application of continuous PEO process was demonstrated on WE43 Mg wires of 0.3 mm in diameter manufactured by cold drawing. Cold drawing is an standard method to manufacture thin metallic wires [23] although the limited reduction of the diameter during cold drawing requires many passes to obtain very thin wires. However, cold drawing outperforms hot drawing and hot extrusion because of the strength of the wires (because of the smaller grain size) and of the possibility to reach very small diameters (< 300 µm) with high precision. The cold drawn wires can be further processed by annealing treatments to tailor mechanical and corrosion properties, which is critical for many applications that require flexible wires. The continuous PEO led to a homogeneous oxide layer whose thickness could be controlled by different parameters, in particular the time of immersion in the electrolyte bath. We also showed that the porosity of the oxide layer could be tailored by changing the electrical parameters of the PEO process. This second part of our work explores the effect of the processing parameters during cold drawing, i.e. amount of cold work and annealing times and temperatures, as well of surface-modification by PEO on degradation rate, deterioration of mechanical properties in simulated body fluid, underlying corrosion mechanisms and cytocompatibility. Our results demonstrate that surface-modification by PEO leads to significant improvements in corrosion resistance and hence biocompatibility of the wires while the reduction of the mechanical properties with immersion time in simulated body fluid is reasonable and predictable. As a result, these factors open up very appealing opportunities for the application of Mg wires in biomedical and, particularly, implant applications.

## 2. Materials and Experimental techniques

### 2.1 Materials

Mg wires of 0.3 mm in diameter were manufactured by cold drawing following the strategy reported in the first part of our work [22]. Two wires (CW34 and CW13) were manufactured to an extent of 34% and 13% cold work. In addition, wires manufactured with 13% of cold work were subjected to four different annealing heat treatments lasting 5 s and 10 s at either 400ºC (HT400-5 and HT400-10) or 450ºC (HT450-5 and HT450-10). Moreover, wires with 13% cold work, followed by annealing at 450ºC for 5 s were additionally surface modified by different continuous PEO treatments. The wire was passed through a phosphate-based electrolyte bath (Kermasorb®, Meotec GmbH, Aachen, Germany) while applying an electrical current of 0.5 A (~15A/dm$^2$) in pulsed DC mode at 330 V and either 250 Hz (PEO250) or 500 Hz (PEO500) in frequency, leading to a continuous process (C-PEO) to convert the surface of the Mg wires and introduce a functional oxide layer. Details of the processing techniques as well as microstructure of the wires and the oxide layers are reported in [22].

### 2.2 In vitro degradation tests

*In vitro* degradation tests of the Mg wires were carried out at 37ºC in 100 mL of simulated body fluid (c-SBF) with a composition (per liter) of 8.035 g NaCl, 0.355 g NaHCO$_3$, 0.225 g KCl, 0.176 g K$_2$(HPO$_4$), 0.145 g MgCl$_2$, 0.072 g Na$_2$SO$_4$ and 50 mL of Tris buffer at pH 7.5 [24]. The ratio of c-SBF volume to the wire surface area was approximately 0.5 mL/mm$^2$ (> 0.2 mL/mm$^2$ as recommended by ASTM G31-72 [25]).

Wires of 280 mm in length with 4 mm sealed at both ends were placed inside the glass burette with the lower end fixed at bottom level of the burette and the burettes were immersed into sealed plastic bottles filled with media, as shown in [24]. The degradation rate was monitored by the generation of hydrogen gas as a function of time, which was normalized with respect to the initial surface area of the wire. Wires to study the effect of immersion in c-SBF on the tensile properties were 40 mm longer at both ends (total length of 280 mm) and sealed with a rubber coating (Liquid rubber coating, Mibenco, Germany) to protect the outward regions from degradation and, hence, to avoid failure in the gripping zones during the tensile tests. At least four wires were tested in tension with electro-mechanical testing machine (ProLine Z010, ZwickRoell GmbH & Co. KG, Ulm, Germany) at a cross-head speed of 5 mm/min. The load was monitored with a load cell of 10 kN.

After the degradation test, random cross-sections of the wires were mounted in an epoxy resin and grounded with 4000 SiC paper. Afterwards, they were polished with 3 μm diamond suspension followed by 0.2 μm $SiO_2$ suspension. Cross-sections were gently cleaned in ethanol. For pitting quantification, random cross-sectional images of degraded HT450-5 (n = 20) and PEO500 (n=8) Mg wires were taken to an optical microscope (VK-X3000, Keyence Deutschland GmbH, Germany). Cross-sections and lateral surfaces were also analyzed in a scanning electron microscope (Zeiss EVO MA15, Germany) after sputtering a thin gold layer at an accelerated voltage of 10~20 kV using secondary electrons as well as energy dispersive X-ray microanalysis (EDS) to study the underlying corrosion mechanisms.

**2.3 Cell culture testing**

The biocompatibility of the wires was tested following ISO 10993-1 and 10993-5. Indirect and direct tests were carried out using pre-osteoblasts from the cell line MC3T3-E1 Subclone 4 (ATCC-CRL-2593). They were grown in a culture medium consisting of alpha MEM medium (Invitrogen, USA), supplemented with 10% of fetal bovine serum (FBS) (Invitrogen,

USA), and 1% penicillin/streptomycin (Invitrogen, USA). Extracts for the indirect tests were obtained by immersion of the Mg wires (6 cm$^2$/mL) in the culture medium, followed by incubation for 24 h at 37°C. 10.000 cells/cm$^2$ were seeded in a 96-well cell culture plate and incubated for 24 h at 37°C in an atmosphere with 5% $CO_2$ concentration and 95% relative humidity. Then, the culture medium was replaced with the concentrated and diluted extracts and incubated for another 24 h and the mitochondrial activity of the cells was measured with the tetrazolium-based MTT assay. For this purpose (3-(4, 5-dimethyl thiazolyl-2)-2, 5-diphenyltetrazolium bromide) MTT (Sigma-Aldrich, Germany) was prepared in a concentration of 5 mg/mL and added to the cells in a proportion of 10 µl of MTT stock solution by 100 µl of culture medium. Cells were incubated for 3 h protected from light. After that, formazan crystals were dissolved by adding dimethyl sulfoxide (DMSO) (Sigma-Aldrich, Germany), and the absorbance of the solutions was measured at 570 nm in a spectrophotometer (Tecan Infinite Mplex). Mitochondrial activity (%) was calculated using the value obtained from the control sample of pure culture medium as a reference. The experiments were performed in triplicates for each sample.

For the direct tests, Mg wires with and without surface modification by PEO were first immersed in culture medium for 24 h to undergo some degradation and then used to study the cell-material interaction. Immersion in the culture medium for 24 h was chosen since differences in the degradation rates between unmodified and PEO surface-modified wires become prominent after this time. Mg wires were mounted on a thin frame of medical grade poly-lactic acid (PLA) (Purasorb 7038, Corbion, Netherlands) obtained from a hot-pressed sheet of PLA of 0.8 mm in thickness. The ends of the wires were glued to the frame by using a PLA/chloroform solution where chloroform was evaporated after attachment. This experimental setup enabled analysis of the Mg wires under the microscope because wires remained in a stable position during handling and, hence, this procedure can serve as a guideline for future research aiming at direct cell material interaction on thin bioresorbable wires. The Mg wires glued to the PLA frame and the Titanium (99% Ti, Alfa Aesar) control were immersed in cell culture medium in a 12-well plate (10,000 cells/cm$^2$) and placed in an incubator for 24 h. Afterwards, the cell-material interaction was analyzed using immunofluorescent microscopy. Cell morphology was then studied by labeling the cytoskeleton of the cells with Alexa fluor 488 phalloidin (Invitrogen, USA) and the nucleus with Dapi (Sigma-Aldrich, Germany). Fixed cells were treated with 0.1% Triton X-100 for 10 min, then cells were washed three times with PBS and labeled with Alexa 488., and Dapi for 1 h. Finally, cells were observed under a confocal microscope (Olympus FV3000, Japan). Cell morphology was also analyzed under scanning electron microscope (SEM) (Zeiss EVO MA15, Germany). For this purpose, fixed cells were washed with PBS and then dehydrated with an increasing concentration of ethanol in PBS (30, 50,70, 80, 90, and 100%) followed by the overnight drying in air. The processed cells were then analyzed using back-scattered electrons at an accelerated voltage of 5-7 kV.

## 3. Results

### 3.1. Corrosion resistance of cold drawn and heat-treated WE43 Mg wires

The early degradation of the Mg wires in c-SBF at 37ºC was assessed from the release of hydrogen gas and the corresponding results for wires subjected to different processing conditions are plotted in Fig. 1. The highest corrosion rate after 1 h of immersion was found for the CW34 wire while the lowest corrosion rate was observed for CW13 wire. Annealing of the CW13 wires led to a slightly increase in the corrosion rate after 1 h but the differences between annealing treatments were negligible. After 6 hours of immersion, the lowest corrosion rate was still found for the CW13 wire. Increasing the amount of cold drawing (CW34) or the annealing temperature or time seemed to enhance the corrosion rate. The effect of annealing on the corrosion rate can be related to the nucleation of precipitates at the grain boundaries, as shown in the transmission electron microscopy observations in [22] while effect of cold work can be linked to the increase in dislocations. It should be noted that the mass loss of the Mg wires after 6 hours of immersion was in the range 20-30% and the differences in the corrosion rates of the Mg wires subjected to various processing conditions were still found to be very limited. Hence, our results clearly indicate that microstructural changes are not enough to tailor the reactivity of Mg alloys and that strategies based on surface modification are necessary.

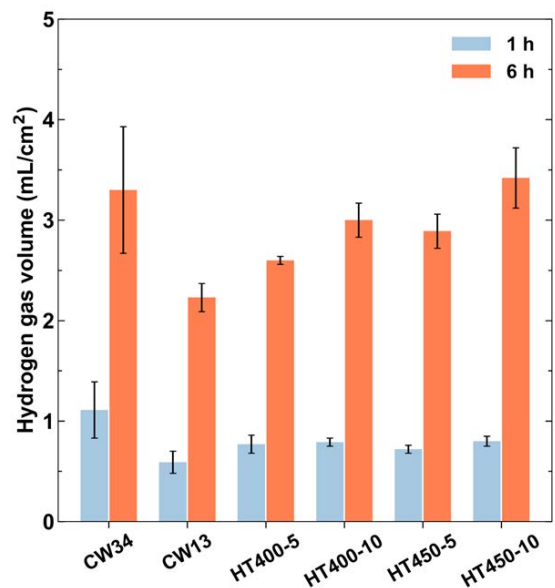

**Fig. 1.** Degradation rate of Mg wires with different processing conditions measured by the hydrogen evolution after 1 and 6 h immersion in c-SBF.

## 3.2. Corrosion resistance and mechanical properties of PEO surface modified Mg wires

Based on the mechanical properties reported in [22] and in degradation properties above, the Mg wire subjected to 13% cold drawing and then annealed at 450ºC for 5 s (HT450-5) was selected for the surface-modification by PEO. The surface treatments were carried out using the parameters as indicated in the materials and methods sections using two different frequencies, namely 250 Hz (PEO250) and 500 Hz (PEO500). The immersion time in the electrolytic bath was always 20 s, leading to a continuous and homogeneous oxide layer of 8.3 ± 1.3 μm and 7.8 ± 1.3 μm for PEO250 and PEO500, respectively. Both oxide layers were porous while apparently PEO250 showed larger pore sizes. Their microstructures are reported in Fig. 10 in the first part of our work [22].

As expected, the continuous oxide layer created by PEO did have a strong effect on the *in vitro* degradation of the Mg wires. The hydrogen volume generated by HT450-5, PEO250 and PEO500 wires as a function of the immersion time in c-SBF at 37ºC is plotted in Fig. 2a. The initial corrosion rate (as measured by the hydrogen evolution) was similar for wires without and with oxide layer during the first ≈ 5 h. However, the hydrogen evolution evolved linearly with time in the unprotected wires up to ≈ 20 h while it was significantly reduced for PEO surface-modified wires after 6 - 7 h.

Likewise, the mechanical properties, i.e. tensile strength, of the degraded wires was measured as a function of the immersion time in c-SBF, see Fig. 2b. While the strength of the HT450-5 wires decreased to almost 0 after 24 h of immersion in c-SBF at 37ºC, the strength of the wires protected by PEO was > 200 MPa after 24 h and > 100 MPa after 96 h. It should be noted that the decrease in mechanical strength of PEO250 and PEO500 wires was very similar up to 96 h, which is in agreement with the results of the corrosion tests. The tensile stress-strain curves in Figs. 2c, 2d and 2e show that the failure strain of all wires decreased with degradation time. This phenomenon was much more marked in the HT450-5 wires which presented severe pitting corrosion. Thus, these wires failed at very low strains at the section with the smallest cross-section in the wire in which deformation was localized. Surface modification of the Mg wires by PEO provided higher strain-to-failure in all cases. Tensile tests could not be carried out on wires immersed for longer times than depicted in the figure since they lost mechanical integrity and were too fragile to test reliably.

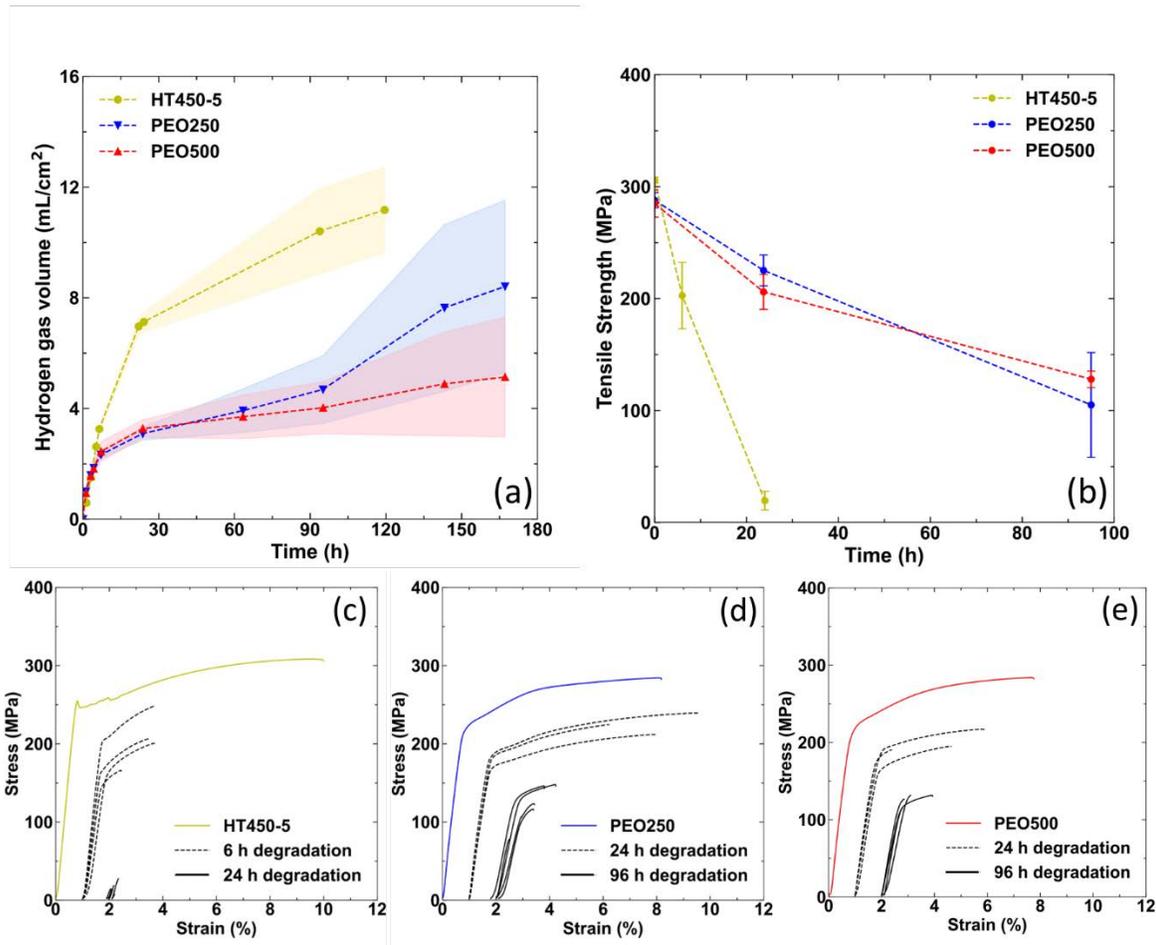

**Fig. 2.** (a) Hydrogen gas evolution as a function of immersion time in c-SBF at 37ºC for Mg wires before (HT450-5) and after (PEO250 and PEO500) surface treatment by PEO. (b) *Idem*

for the tensile strength. The lower row shows changes in the tensile stress-strain curves of (c) HT450-5, (d) PEO250 and (e) PE0500 wires as a function of degradation time. Shaded regions in (a) indicate the standard deviation.

The fracture region of the PEO500 wire tested in tension before immersion in c-SBF is depicted in Fig. 3a. The fracture surface is oriented at 45º from the tensile loading axis, indicating that the final fracture process took place by shear, a typical mechanism in Mg alloys. The image at higher magnification shows many cracks, perpendicular to the loading axis, in the oxide layer. They indicate that the oxide layer was well adhered to the Mg wire and failed in tension because of the load transfer during the plastic deformation of the wire. The fracture region of the PEO500 wire deformed after 96 h of immersion in c-SBF is depicted in Fig. 3b. It shows a core region of Mg, which also failed by shear, surrounded by a thick circumferential oxide layer, which is homogenously distributed and did not detach from the Mg core. On the contrary, the fractures of Mg wires that were not modified by PEO showed to be fully brittle after 24 h degradation (Fig. 3c) and the strain to failure was negligible.

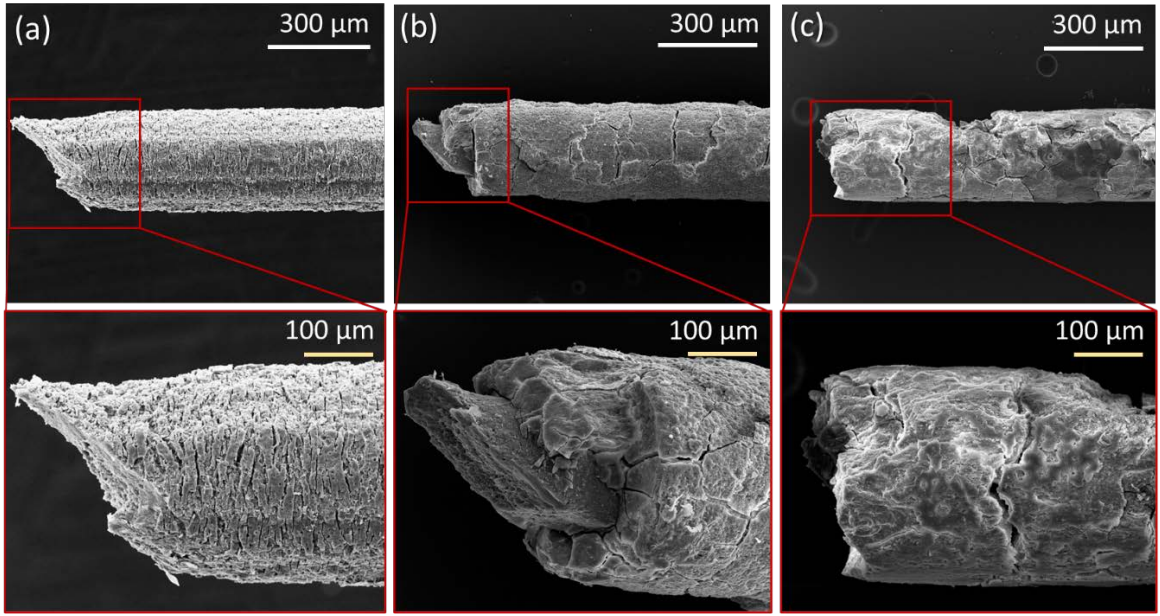

**Fig. 3.** SEM images of Mg wires before and after degradation. (a) Fracture region of the PEO500 wire deformed under tension before immersion in c-SBF. (b) *Idem* after 96 h in c-SBF. (c) Fraction region of a HT450-5 wire after 24 h immersion in c-SBF.

### 3.3. Corrosion mode of Mg wires without and with surface modification by PEO.

The presence of the oxide layer created by PEO did not only reduce the average corrosion rate of the wires but also reduced the intensity of pitting corrosion. This is shown in Figs. 4a and 4b, in which random cross-sections of the Mg wires (including only the Mg section of the wire) are depicted for the HT450-5 wire after 24 h of immersion in c-SBF and for the PEO500 wire after 96 h of immersion in c-SBF, respectively. For the wire without any surface protection by PEO, corrosion was easy to localize in random cross-sections of the wire and the mechanical strength of the wires was quickly reduced to a negligible value within 24 h of immersion. On the contrary, the PEO oxide layer showed to hinder the development of severe pitting corrosion and the Mg cross-section of these wires remained almost circular and intact even after 96 h of immersion in c-SBF.

The differences in corrosion profile were quantified by evaluating the severity of pitting corrosion by image processing using PitScan [24]. To this end, cross-sectional images of degraded wires were analyzed following the procedure detailed in the supplementary information. The metrics that characterize the intensity of pitting corrosion are summarized in Fig. 5. They are the uniform corrosion radius in Fig. 5a (the radius of the circular section that has the same area as the corroded cross-section), the average pit depth in Fig. 5b (average distance from the degraded cross-section to the uniform corrosion circle), and the maximum pit depth in Fig. 5c (maximum distance from the corroded cross-section to the uniform corrosion circle). Fig. 5a confirms the hydrogen evolution results reported in Fig. 2a and the uniform corrosion radius of PEO500 (113 ± 8.5 µm) was still larger after 96 h than that of the non-surface treated HT450-5 (108 ± 21 µm) after only 24 h. Moreover, the high standard deviation of the uniform corrosion radius of HT450-5 shows the greater variation of mass loss at different cross sections with respect to PEO500. The differences in the intensity of pitting corrosion between both wires also become evident in the average pit depth and in the maximum pit depth as plotted in Figs. 5b and 5c. The reported localization of damage was responsible for the obvious drop in strength and ductility of the HT450-5 wires when immersed in c-SBF.

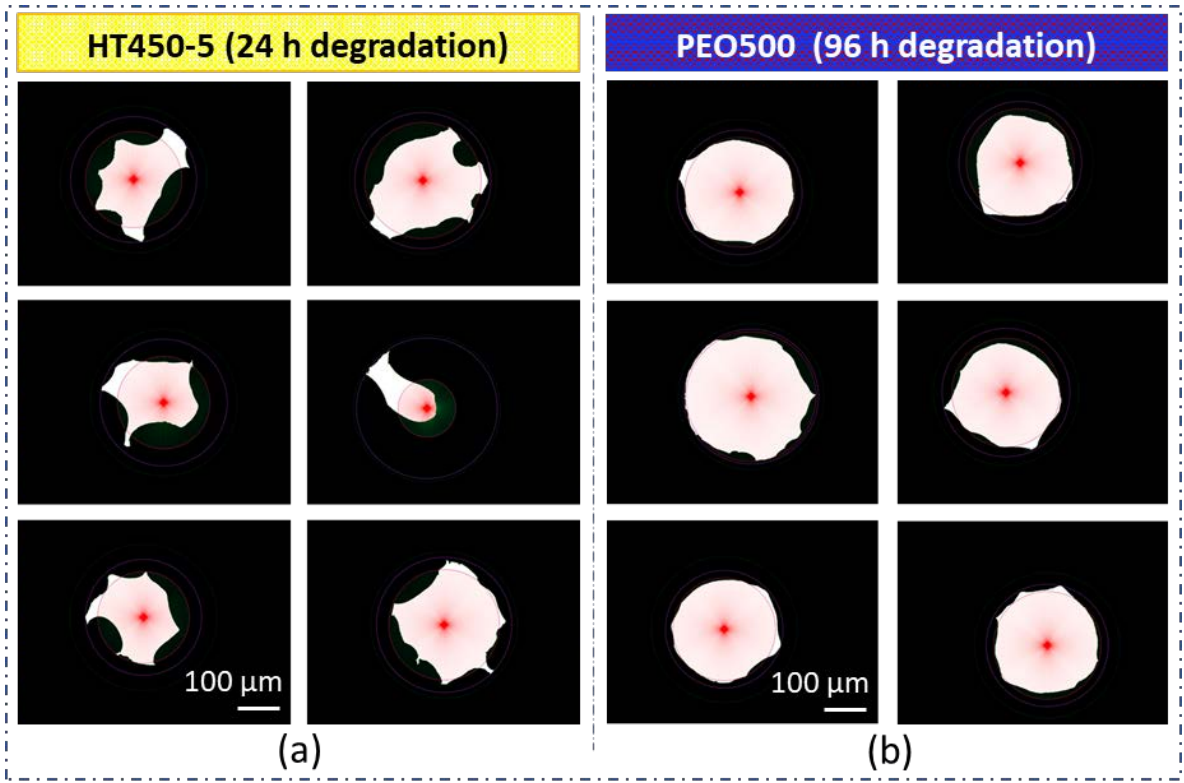

**Fig. 4.** (a) Representative cross-sections of HT450-5 wire after 96 h immersion in c-SBF. (b) Idem of PEO500 wire after 96 h immersion in c-SBF. Red points mark the original center of cross section before degradation. The scale bar of 100 μm is for all cross-sections in this figure.

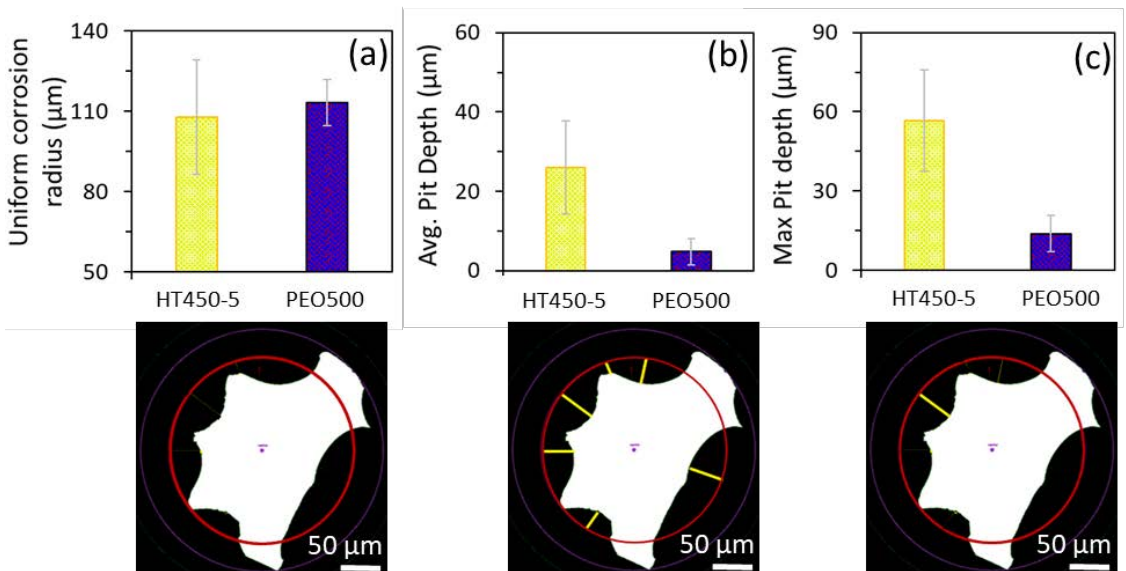

**Fig. 5.** Metrics representing the intensity of pitting corrosion in HT450-5 and PEO500 wires and representative cross sectionns of degraded HT450-5 Mg wire: (a) Uniform

corrosion radius (marked by red circle). (b) Average pit depth (each yellow line marks the deepest point of each pit). (c) Maximum pitting depth (marked by yellow line).

### 3.4. Biological performance of PEO surface modified Mg wires

The mitochondrial activity of the MC3T3-E1 preosteoblasts after 24 h of incubation in extracts obtained from different Mg wires (HT450-5, PEO250, PEO500) immersed in the culture medium is depicted in Fig. 6a over different dilutions. Pure extracts equal 100% of value while 50%, 25%, 12.5% and 6.25% indicate the concentration of the pure extract in culture medium. HT450-5 Mg wire showed slight toxic behavior in 100% extract with a cell viability of 68.4 ± 5.5 %, while both PEO-modified Mg wires did not show any cytotoxic potential. These results are supported by the optical microscopy images of 100% extracts in Fig. 6b, 6c and 6d which show alive (elongated) and dead (circular, marked with an arrow) cells in the 100% extracts of HT450-5, PEO250 and PEO500 wires, respectively, after 24 h of incubation. While the results in Fig. 6a may be biased by the presence of corrosion products that can change the absorbance of the solutions, the fraction of dead cells in the 100% extracts followed the order HT450-5 > PEO250 > PEO500 and shows the slightly elevated cytotoxic activity of the HT450-5 wires due to more pronounced degradation and associated effects, such as pH increase and higher hydrogen gas evolution. Also, MC3T3-E1 cell viability is known to drop when the concentration of Mg ions in the culture medium increases [26], which is in agreement with these observations. However, no sample showed any toxicity already after the first dilution.

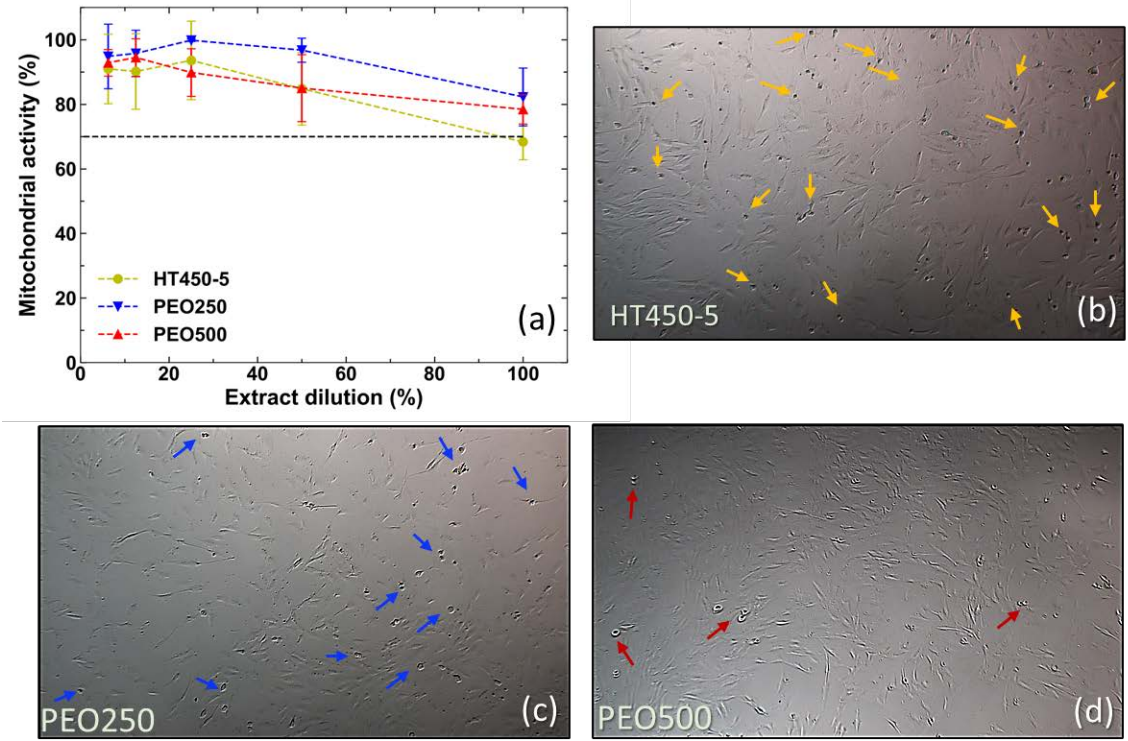

**Fig. 6.** (a) Mytochondrial activity of the MC3T3-E1 pre-osteoblasts after 24 h of incubation in extracts obtained from Mg wires immersed in culture medium. (b), (c) and (d) Optical micrographs of living (elongated) and dead (circular, marked with arrows) cells in the 100 % extracts of HT450-5, PEO250 and PEO500 Mg wires after 24 hours. Magnification was 10x. Arrows point to dead cells which show a dark round shape.

The direct interaction between the cells and the Mg wires was assessed by means of observations under a confocal microscope and SEM (Fig. ). After 24 h, the MC3T3-E1 cells covered the surface of the Ti control samples by cell-cell and cell-material interactions, as shown by both confocal and SEM images in Fig. 7a. On the contrary, cells could not attach on the surface of the HT450-5 Mg wire (Fig. 7b) which can be linked with the dynamic environment on the metallic surface of the wires in which fast corrosion leads to hydrogen gas release and/or to a local alkaline environment. The surface of the HT450-5 Mg wire appeared fully cracked in the SEM because of the dehydration and shrinkage of the corrosion layer by dehydration already in air or under the vacuum of the SEM. On the contrary, many living cells were able to attach on the surface of the PEO250 and PEO500 Mg wires. The connection between cells and the PEO surface was shown in both cases by bridging filopodia structure which generally is evidence of a good cell adhesion to surface [18][27][28]. Qualitatively both PEO250 and PEO500 Mg wires had same favorable cell attachment behavior. This behavior indicates that the presence of the oxide layer was necessary to enable a proper adhesion of cells to the wire surface by providing a porous surface with stable environment in terms of gas release and surface alkalinity.

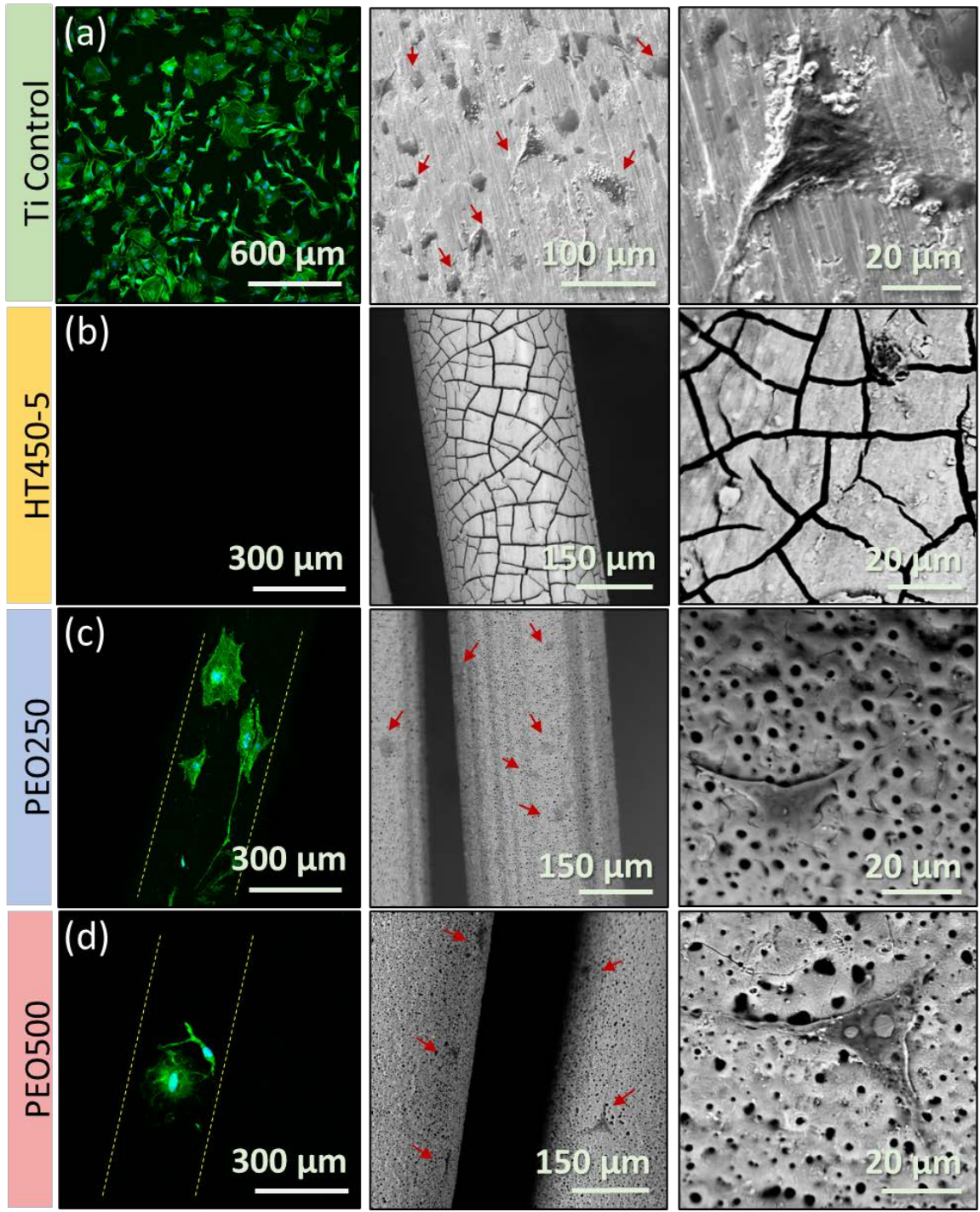

**Fig. 7.** Confocal microscopy and back-scattered SEM images of the surface interaction between MC3T3 pre-osteoblasts and the tested Mg wires for 24 h. (a) Ti control sample. (b) HT450-5. (c) PEO250. (d) PEO500. Yellow dashed lines indicate the wire edges. Red arrows indicate the presence of cells on the wire surface.

## 4. Discussion

### 4.1 Corrosion of WE43 Mg wires

The corrosion resistance of WE43 Mg wires manufactured by cold drawing and subjected to different annealing treatments were measured by hydrogen evolution tests in c-SBF at 37ºC and the mechanical properties of the wires were measured as a function of the immersion time. Our results show that the corrosion resistance of the wires was affected by the amount of cold work and the annealing treatments, but only to a minor extent. The highest corrosion resistance was obtained on wires subjected to the lowest amount of cold work (CW13) without any annealing. More cold work or annealing increased the corrosion rate in agreement with previous investigations [11] [12] that, however, did not yet provide a conclusive explanation. In our study, the effect of annealing on the corrosion rate could be associated with the appearance of nm-sized precipitates at the grain boundaries [22]. They are known to increase the corrosion rate due to the activation of galvanic corrosion mechanisms [29][30] which is even more pronounced in materials with small grain size. Similarly, dislocation pileups near the grain boundaries lead to structures that are far from equilibrium and that can also be susceptible to corrosion. Nevertheless, the corrosion rates of Mg wires are always very high regardless of the amount of cold work and annealing treatments and mass losses in the range 20% to 30% are already achieved within 6 h of *in vitro* tests because of the high specific surface. On top of that, pitting is very active Mg wires, leading to the localization of corrosion in particular sections of the wire and to a complete loss of mechanical properties within 24 h of immersion in c-SBF at 37ºC. It is unlikely that improvements in corrosion resistance by alloying and/or thermo-mechanical treatments offer the potential to substantially change this phenomena [11] [31]. Moreover, although our indirect cytocompatibility tests only showed slight toxic levels (68.4% of cell viability with respect to the control samples), direct tests did not show any MC3T3 preosteoblast cell attachment on the surface of the Mg wires that were not modified by PEO. Similar results have been reported in porous Mg scaffolds manufactured by laser power-bed fusion, which also present a very high specific surface, if their surface was not modified by PEO [29]. Thus, (electro-chemical) passivation of the surface of Mg wires seems to be a must to reduce the corrosion rate, maintain mechanical properties after immersion in an aqueous environment (e.g. c-SBF) and provide surfaces that are suitable for cell adhesion and proliferation in medical use.

**4.2 Corrosion of WE43 Mg wires surface-modified by PEO**

The continuous PEO process presented in the first part of our work [22] showed successful to grow an homogenous and porous oxide layer on Mg wires and is easily scalable to manufacture wires by combination of cold drawing, heat treatment processes and surface treatment. The corrosion mechanisms of the Mg wires with and without surface-modification by PEO were elucidated through the analysis of the cross-sections of the wires using SEM and EDX. The respective images and mappings for HT4505 wire after 24 h of immersion in c-SBF at 37ºC are plotted in Fig. 8, and they confirm the presence of Mg, O, P and Ca in the corrosion layer. The cracks in the corrosion layer were induced by dehydration of samples. The interaction of c-SBF with Mg surfaces is well established in literature [32]. The water molecules in the c-SBF rapidly react with the Mg on the wire surface, forming a layer of $Mg(OH)_2$ and releasing $H_2$ gas. This layer is not stable in the presence of $Cl^-$ ions, leading to the formation of soluble $MgCl_2$ and $OH^-$ ions. As a result of the deterioration of the $Mg(OH)_2$ layer, Mg is exposed in the surface and the corrosion process progresses towards the inside of the wire by the successive formation (Eq. 1) and dissolution (Eq. 2) of Mg(OH)$_2$ according to the following chemical equations:

$$Mg + 2H_2O \rightarrow Mg(OH)_2 + H_2 \qquad (1)$$

$$Mg(OH)_2 + Cl^- \rightarrow MgCl_2 + 2OH^- \qquad (2)$$

$$HCO_3^- + OH^- \rightarrow H_2O + CO_3^{2-} \qquad (3)$$

$$CaCl_2 \rightarrow Ca^{2+} + 2Cl^- \qquad (4)$$

$$Mg^{2+} + CO_3^{2-} \rightarrow Magnesium\ Carbonates \qquad (5)$$

$$Ca^{2+} + (HPO_4^{2-}, CO_3^{2-}) + H_2O \rightarrow Calcium\ Phophates\ and\ Carbonates \qquad (6)$$

Even though the corrosion layer becomes thicker during the corrosion progress, it is not compact enough to stop the corrosion by diffusion of the c-SBF solution. The amount of P and Ca decreases from the outer edge of the corrosion layer to the inner core of the Mg (Fig. 8) suggesting the formation of Ca-P phosphates and carbonates following the reactions indicated above (Eqs. 3, 4, 5 and 6), in agreement with a recent report [33]. The reduction in Ca and P concentration with depth is likely due to the slow transport of Ca$^{2+}$ ions through degradation layer, thus promoting its precipitation near the outer edge. The exact composition of Calcium Phosphates depends on the Ca/P ratio [34].

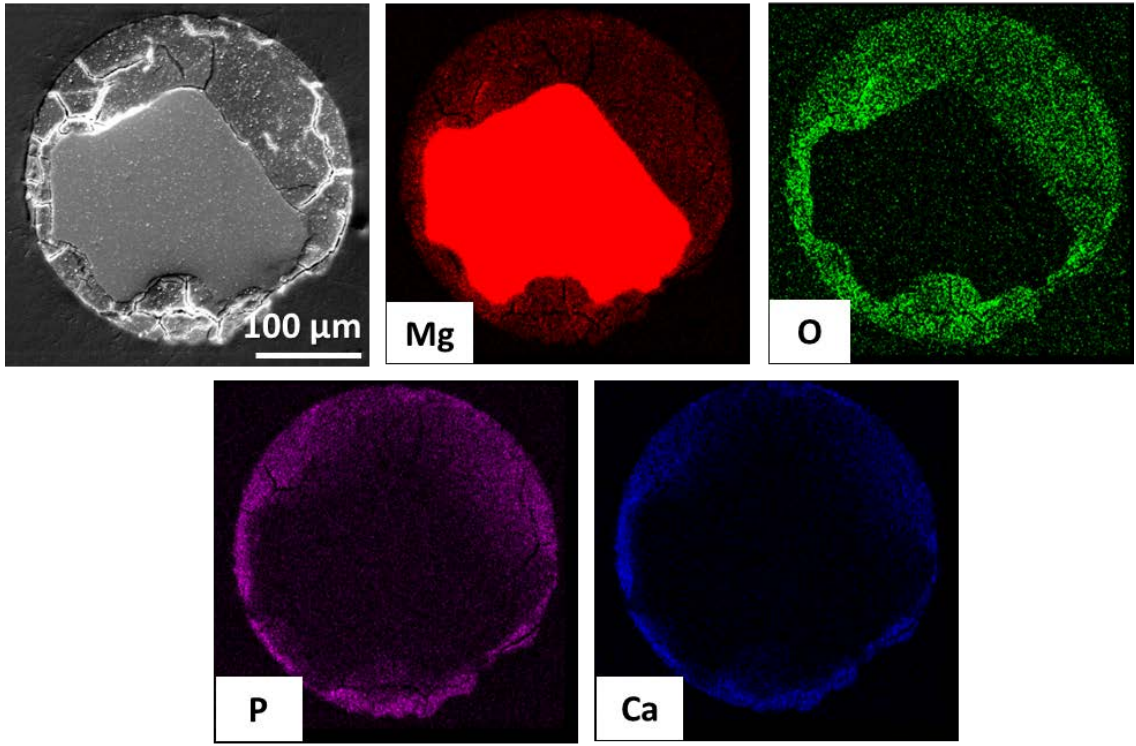

**Fig. 8.** Analysis of the cross-section of HT450-5 wire under SEM/EDS after 24 h of immersion in c-SBF at 37ºC. The first image is obtained with secondary electrons and the other images stand for the elemental composition maps representing Mg, O, P and Ca were obtained by EDX.

The corrosion mechanisms in the presence of an oxide layer created by PEO seem very different, as shown in the cross-sections of the PE500 wire before (Fig. 9a), after 96 h (Fig. 9b) and after 168 h (Fig. 9c) of degradation in c-SBF at 37ºC. The chemical composition of the continuous oxide layer before degradation includes Mg, O and P and, thus, consists mainly of MgO and $Mg_3(PO_4)$ [22]. Independent of the closed barrier layer covering the immediate interface to the metal substrate, the PEO oxide layer has an outwards open porous surface. Hence, aqueous molecules from the c-SBF solution might be able to reach the surface of the Mg wire after breaking the barrier layer by corrosive attack, leading to the same corrosion mechanisms described above for HT450-5 Mg wires. Thus, similar values for the initial corrosion rates of both HT450-5 and PEO500 wires might by justified. However, after a certain time in the corrosive environment, the degradation rate of the PEO surface-modified wire is significantly reduced which might be associated with the appearance of an external protective and dense corrosion layer containing O, Ca and P (Fig. 9b). This dense layer is located beneath the PEO but on top of the typical porous oxide layer resulting from the progressive oxidation of Mg to form $Mg(OH)_2$, which is found between the external Ca-P dense layer and the Mg core of the wire. Even not fully inhibiting diffusion processes, the presence of this dense Ca-P layers seems to slow down migration of $Cl^-$ and other corrosive ions which in turn decreases the elimination of the $Mg(OH)_2$ layer (Eq. 2). As a result, the corrosion rate of the Mg wire is reduced, and pitting corrosion is suppressed. This protective process seems to continue over degradation time and the dense and continuous Ca-P layer is still seen on top the Mg oxide layer even after 168 h of immersion in c-SBF (Fig. 9c) while the remaining Mg core is still present in the center with unaltered oval with no prominent pits.

As a result, we attribute the reduced corrosion rate and the uniform degradation on PEO surface-modified WE43 Mg wires to the presence of this Ca-P rich layer deposited during initial stages of the corrosion process. Even though a Ca-P layer growing over time was also observed for the HT450-5 wires, this natural corrosion layer was not as compact and continuous as for the PEO variants. Since there is no evidence in literature about the formation of a dense and continuous Ca-P layer during corrosion of PEO surface modified Mg, the following explanation for its appearance is hypothesized. During the first stages of corrosion, the initial oxide layer created by PEO on top of the Mg wire is cracked because of the expansion associated with corrosion. This oxide layer contains $Mg_3(PO_4)_2$ and might supply sufficient phosphate ions to react with the $Ca^{2+}$ ions derived from the c-SBF, leading to the precipitation of an additional Ca-P rich layer on top of the $Mg(OH)_2$ which is formed during corrosion (Fig. 10). The equal intensity of P signal (Fig. 10) in the oxide layer created by PEO and in the Ca-P layer also supports the above argument while the presence of Ca at the inner location of the PEO oxide layer is also in agreement with this assumption.

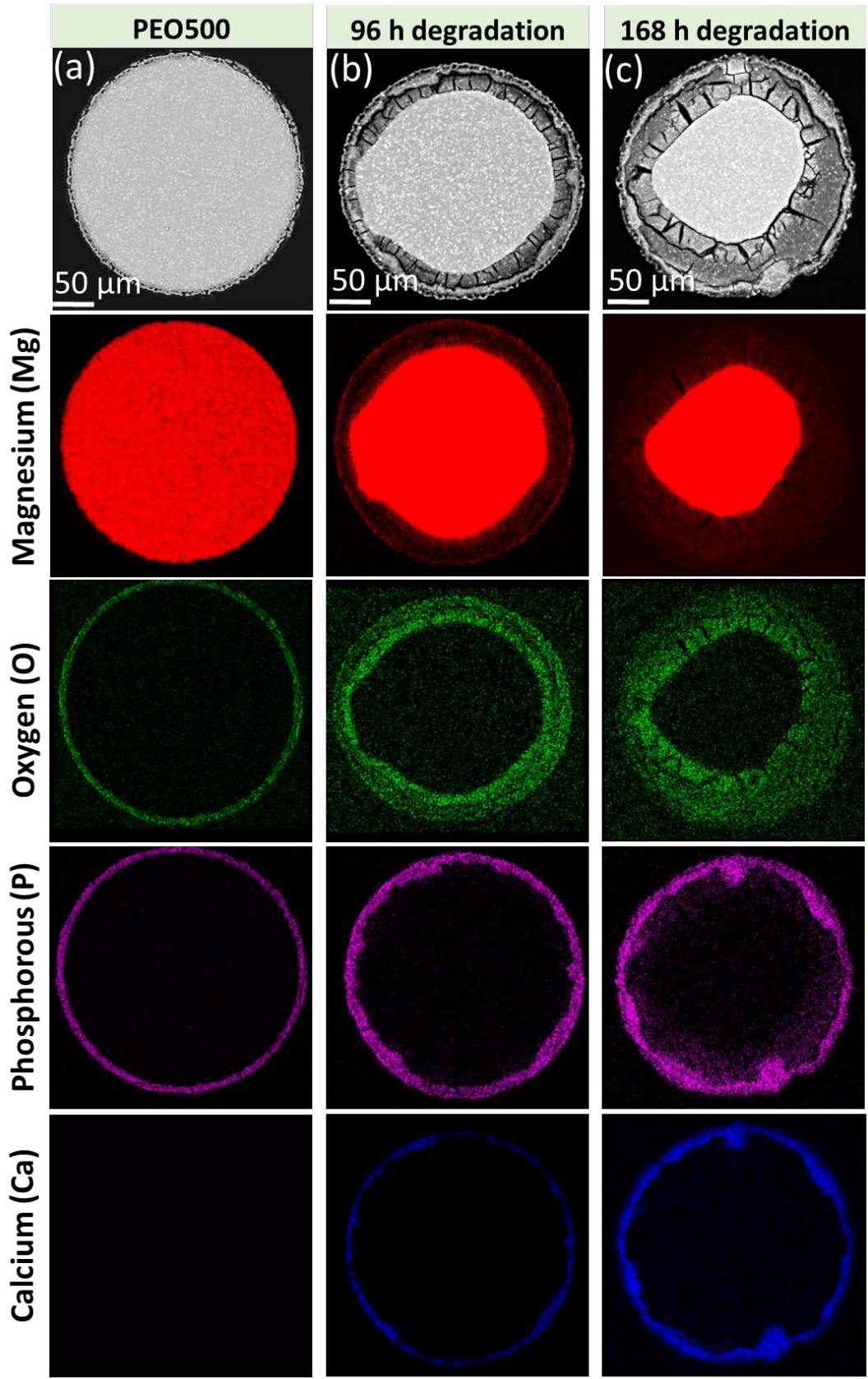
26

**Fig. 9.** Analysis of the cross-section of the PEO wire (a) before and after (b) 96 h and (c) 168 h of degradation in c-SBF at 37ºC. The first column shows the back-scattered electron image and the other columns represent the compositional distribution of Mg, O, P and Ca obtained by EDX.

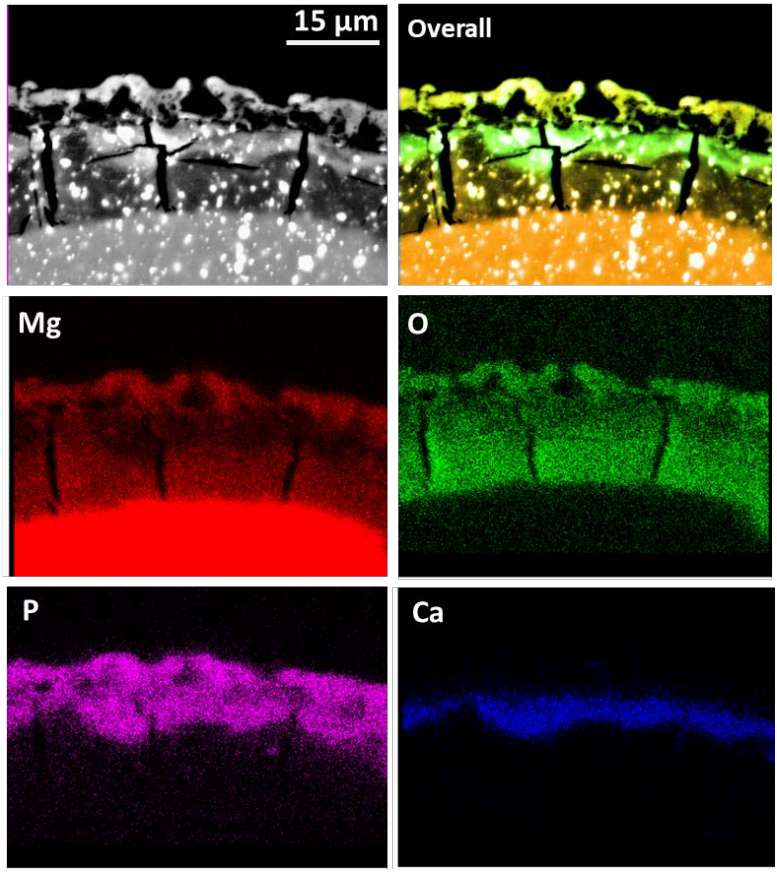

**Fig. 10.** Higher magnification analysis of the surface of the PEO500 wire after 96 h in SBF. The first image on the top left shows the back-scattered electron image and the other present the compositional distribution of Mg, O, P and Ca obtained by EDX.

After degradation for 96 h and 196 h in c-SBF, cracks were found in particular on the lateral surfaces of the PEO250 (Figs. 11a and 11b) and PEO500 wires (Figs. 11c and 11d). These cracks seemed to be formed because of the internal stress induced by the expansion of corrosion products [35] and to facilitate the influx of the media to the Mg core, leading to the partial detachment of the external corrosion layer. However, the degree of surface degradation seemed to rely on the nature of the initial PEO oxide layer since it was much more pronounced in the PEO250 wire, in which the external corrosion layer completely detached after 168 h in the c-SBF (Fig. 11b). On the contrary, the external oxide layer was mostly intact (Fig. 11d) for the PEO500 wire although cracks were also found on the surface of this variant. These results confirm the importance of optimizing electrical parameter of the PEO process to slow the degradation of the wires and to promote the formation of the Ca-P dense layer upon degradation of Mg core. Moreover, they also explain why the corrosion rate of both PEO surface-modified wires was similar up to 96 h after which the corrosion rate of the PEO250 wire apparently increased due the detachment and degradation of the PEO oxide layer.

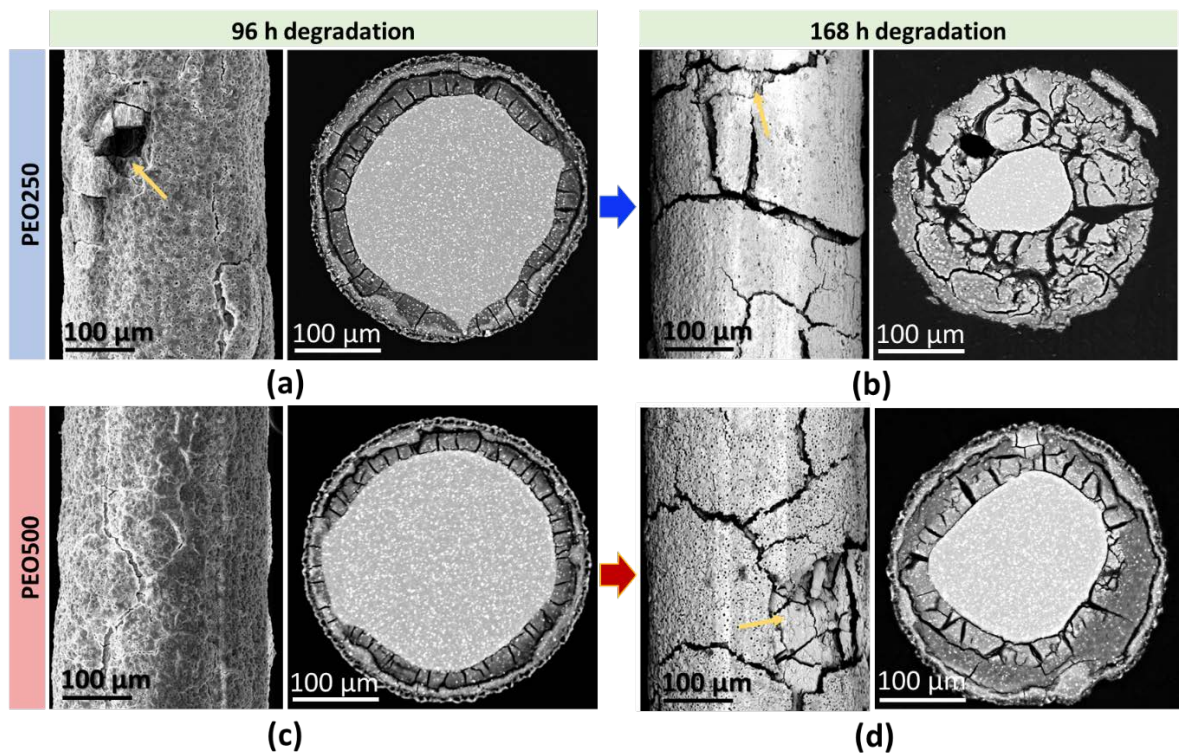

**Fig. 11**. Back-scattered electron images of the cross section and lateral surfaces of PEO250 (a-b) and PEO500 (c-d) Mg wires after 96 h and 168 h of degradation in c-SBF at 37ºC, respectively. Yellow arrows indicate detachment of the PEO oxide layer.

**4.3 Limitations of oxide-layer formed by PEO**

It should be finally noted that the continuous oxide layer promoted by continuous PEO in our work presented good adhesion on the Mg wire but showed some typical brittleness which might lead to fine cracks in the presence of sufficient strains. For instance, multiple parallel cracks appear on the surface when the wires are twisted or knotted, even though no detachment or delamination can be seen even at elevated strains (Fig. 12). To minimize this phenomenon of strain-induced cracking, we propose to reduce the oxide layer thickness (in order of 1~2 microns) which is possible by finely controlling the electrical parameters of the electrochemical PEO process. Another possibility might be to carry out PEO while subjecting the wires to tensile stress, so that the PEO oxide layer will be under compression once the wire is relaxed. These strategies are part of our future work.

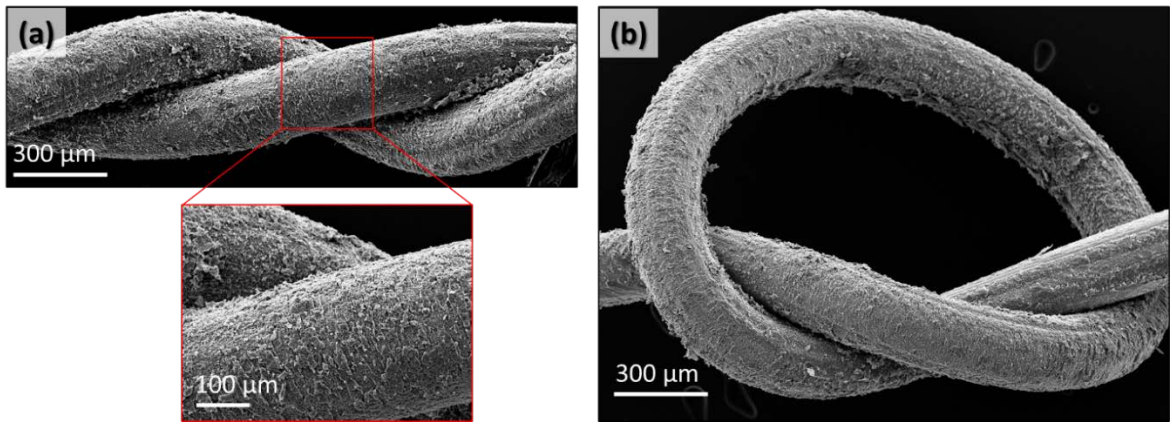

**Fig. 12.** SEM images of (a) twisted and (b) knotted PEO500 Mg wires, showing a fine network of cracks on the PEO oxide layer upon deformation.

**5. Conclusions**

The corrosion resistance and biological performance of cold-drawn WE43 Mg wires were analyzed as a function of its thermo-mechanical processing history and the continuous treatment of the wires to form a plasma-electrolytic oxide layer (C-PEO). It was found that annealing led to slight increase in the corrosion rate due to the nucleation of precipitates at grain boundaries but – in the absence of a protective oxide layer such as created by PEO – the corrosion rate of the Mg wires was rather fast, with dominant pitting corrosion, leading to a decrease in mechanical strength and ductility of the Mg wires already after 24 h of immersion in simulated body fluid at 37ºC. In addition, indirect and direct tests with MC3T3 preosteoblasts showed diminished levels of activity and almost no cell attachment on the Mg wires that were not modified by PEO. We conclude that it seems generally unlikely that any further modification of the microstructure of the wires could be able to change significantly this rash decline in properties as necessary for use as an implant material.

Thus, surface modification of the annealed WE43 Mg wires by a continuous PEO process was developed and led to the creation of a homogeneous protective oxide layer of ≈ 8 μm in thickness comprising of MgO and $Mg_3(PO_4)_2$. This protective layer significantly improved the corrosion resistance and biocompatibility of PEO surface-modified Mg wires. It was found that a dense layer of Ca/P was deposited at the early stages of corrosion within the inner section of the PEO modification and on top of the inner $Mg(OH)_2$ corrosion layer. This dense Ca/P efficiently slowed down the diffusion of the $Cl^-$ ions and other aggressive ions that dissolve $Mg(OH)_2$ and accelerate the corrosion of Mg alloys. As a result, pitting corrosion was significantly suppressed and the strength of the Mg wires could be maintained up to 100 MPa after 96 h of immersion in simulated body fluid at 37ºC which for the first time gives potential for use in biomedical devices. With regards to biological performance, living cells were able to attach on the surface of the PEO surface-modified, Moreover, direct cell culture tests indicated that the PEO oxide layer was favored the adhesion of cells to the wire surface by providing a porous surface with stable environment in terms of gas release and surface alkalinity. Our results demonstrate the potential of Mg wires continuously surface-modified by C-PEO in terms of mechanical, degradation and biological performance ,paving the way for bioabsorbable metallic wire-based medical devices.

## 6. Acknowledgements

This investigation was supported by the European Union's Horizon 2020 research and innovation program under the European Training Network BioImpant (Development of improved bioresorbable materials for orthopaedic and vascular implant applications), Marie Skłodowska-Curie grant agreement No 813869. Additional support from the Spanish Research Agency through the grant PID2021-124389OB-C21 is also gratefully acknowledged. MER acknowledges the support of the Spanish Ministry of Science and Innovation through the Juan de la Cierva-Formation fellowship FJC2019-039925-I.

**Conflict of interests**

Kopp is employed by Meotec GmbH. Presented magnesium wires are in development and commercially not available. The authors declare no conflict of interests.